\documentclass[12pt]{article}
\usepackage{amsmath}
\usepackage{amssymb}
\usepackage[english]{babel}
\usepackage{graphicx}
\usepackage{indentfirst}
\usepackage{hep}
\usepackage{mathrsfs}
\usepackage{dsfont}
\usepackage{setspace}
\usepackage[labelfont=bf,labelsep=period]{caption}

\usepackage[numbers,sort&compress]{natbib}

\newtheorem{proposition}{Proposition}

\newtheorem{corollary}{Corollary}

\begin{document}\baselineskip24pt

\title{Test of Nonlocal Hidden Variable Theory by the Leggett Inequality in High Energy Physics}

\author
{Abdul Sattar Khan,$^{1}$ Jun-Li Li,$^{2,3}$ and Cong-Feng Qiao$^{1,3,*}$ \\ [0.2cm]
\footnotesize{$^{1}$School of Physical Sciences, University of Chinese Academy of Sciences, Beijing 100049, China}\\
\footnotesize{$^{2}$Center of Materials Science and Optoelectronics Engineering \& CMSOT,} \\
\footnotesize{University of Chinese Academy of Sciences, Beijing 100049, China}\\
\footnotesize{$^{3}$Key Laboratory of Vacuum Physics, University of Chinese Academy of Sciences} \\
\footnotesize{Beijing 100049, China}\\
\normalsize{$^\ast$ To whom correspondence should be addressed; E-mail: qiaocf@ucas.ac.cn.}
}

\date{}

\maketitle

\begin{abstract}\doublespacing
The Leggett inequality is a constraint on the bipartite correlation that admits certain types of non-localities. Existing tests mainly focused on the electromagnetic systems where measurement apparatus are assumed to be projective and sharp. However, in nature there are interactions that do not obey the same conservation laws for photon, and the actual measurements may subject to unavoidable uncertainties due to the fundamental physical principles. In this work, we generalize the Leggett inequality to incorporate the measurements that are unsharp and/or biased. It is found that the parity violation in nature provides a spontaneous implementation of an unsharp measurement for the spin of hyperon. A fine structured Leggett inequality for hyperon decays characterized by the asymmetry parameters is obtained and its violation is found which could be observed with the yet obtained data in experiment, like BESIII and Belle.
\end{abstract}

\newpage

\section{Introduction}

The non-locality is a distinct feature of quantum mechanics (QM) and is exhibited in the violations of various Bell inequalities \cite{Bell-1, CHSH-1, CH-1}, which have been verified in photons \cite{Bell-photons}, atoms \cite{Atom-Be}, and hybrid systems \cite{Bell-review}. Most of the experiments, mainly in the electromagnetic realm, favor the quantum predictions and render the joint assumption of realism and locality in Bell inequality untenable. Meanwhile, the attempt to test the Bell inequality in high energy physics, namely, with massive quanta and different interactions, is a long lasting aspiration lingering in physicists' minds. However, the task suffers from some hard-to-surmount issues \cite{Bertlmann-1}, of which a major one is the loss of ``freewill'' when experimenters try to steer the analyzer, or its analog, due to the spontaneous decays of particles \cite{Passive-1}. This is known as the active measurement problem \cite{Passive-2}; see Ref. \cite{Leggett-Baryon-1} for a recent discussion on the entangled baryons and a detailed analysis for the spin-spin correlation with estimations of necessary event number from charmonium decays \cite{hyperonic-B}.

In 2003, Leggett introduced a class of nonlocal model, i.e., relax the requirement of locality while still keep the realism \cite{Leggett-2003}, and formulated an incompatible theorem between the nonlocal realism and QM in terms of the Leggett inequality. Soon after, experiments with photon were performed and shown in conflict with the Leggett model and agree with the quantum predictions \cite{Leggett-Exp1, Leggett-Exp2, Leggett-Exp3, Leggett-NP}. In recently, the falsifications of Leggett model using neutron matter wave and with solid state spins were carried out \cite{Leggett-neutron, Leggett-NVC}. Unlike the Bell inequality, there is no explicit requirement for active measurement in obtaining the Leggett inequality. This enables the experimental test of the nonlocal realism theory to be realized in high energy process beyond the electromagnetic interaction.

The measurements in the experimental test of the Bell inequality and Leggett inequality are assumed to be projective ones which are unbiased and sharp. Though the projective measurement belongs to the positive operator-valued measure (POVM) measurement \cite{QPQI-Book}, in general the POVMs outperform the projective measurement for many quantum information tasks, such as state estimation \cite{Q-alg}, quantum cryptography \cite{Q-crypt}, device-independent randomness certification \cite{Q-random}, etc. The Bell inequalities applicable to the POVM measurements were yet established  \cite{CH-general, CHSH-unsharp-1}; however, how the Leggett inequalities are violated under the general POVM measurement is still unknown.

Here, we generalize the Leggett inequality to incorporate the general POVM measurement which appears biased and unsharp, and show for the first time that the hyperon hadronic decay actually performs an excellent POVM measurement on spin with unsharp outcomes. It should be noted, while the Bell inequalities designated for the POVM can be violated by the unsharp measurement for entangled hyperon pair, the violation of Leggett inequality is found which depends on the asymmetry parameter of hadron decay, that means the POVM measurement does not always guarantee the violation happen. Nevertheless, in this paper, a fine structured Leggett inequality for hyperon decays characterized by the asymmetry parameters is obtained and its violation is found which can be definitely observed in the sufficiently sharp measurement process, i.e., $\eta_c(\chi_{c0})\to \Sigma^+ {\bar{\Sigma}}^- \to (p\pi^0)(\bar{p}\pi^0)$.

\section{The POVM measurement and non-locality}

The POVM operators for qubit (or spin-1/2) system may be defined as
\begin{align}
\mathcal{M}_{+}(\vec{n}) \equiv \frac{\eta^{(+)} +  \alpha \vec{\sigma} \cdot \vec{n}}{2} \; , \; \mathcal{M}_{-}(\vec{n}) \equiv \frac{\eta^{(-)} -  \alpha \vec{\sigma} \cdot \vec{n}}{2} \; . \label{M-eta-alpha}
\end{align}
Here $\vec{\sigma} = (\sigma_{x},\sigma_{y},\sigma_z)$ is an array of the Pauli matrices; $\eta^{(\pm)}$ and $\alpha$ describe the degrees of bias and unsharpness, respectively, with $\eta^{(\pm)} = 1\pm\eta$ and $|\eta\pm\alpha|\leq 1$. Equation (\ref{M-eta-alpha}) will reduce to the projective measurement when the outcome is sharp ($|\alpha|=1$) and has no bias ($\eta = 0$). Taking the outcome ``$+1$'' as an example, the probability for observing $+1$ while performing the measurement along $\vec{n}$ is
\begin{align}
P_{+}(\vec{n}) \equiv \langle\psi |\mathcal{M}_{+}(\vec{n}) |\psi\rangle \; , \label{Probability-M}
\end{align}
which gives $\frac{1+ \eta -|\alpha|}{2} \leq P_{+}(\vec{n}) \leq \frac{1+\eta +|\alpha|}{2}$ for arbitrary quantum state $|\psi\rangle$. If the particle $A$ spins along the direction $\vec{u}$, the average value for measuring the polarization along a direction $\vec{a}$ gives
\begin{align}
\bar{A}_{\vec{u}}(\vec{a}) & = \langle \vec{u}| \mathcal{M}_+(\vec{a}) |\vec{u}\rangle -  \langle \vec{u} | \mathcal{M}_-(\vec{a}) |\vec{u}\rangle \nonumber \\
& = \eta+\alpha \vec{u}\cdot \vec{a} \; . \label{avarange-Meth}
\end{align}
Here $\vec{a}$ is a unit vector and $|\vec{u}\, \rangle$ denotes the quantum state with spin along $\vec{u}$. Obviously, for unbias and sharp  measurements, we will get the Malus' law $\bar{A}_{\vec{u}} (\vec{a}) = \vec{u}\cdot \vec{a}$, the well-known cosine dependence of the intensity of polarized beam through an ideal polarizer.

\subsection{Bell and Leggett inequalities under POVM measurement}

In talking about the hidden variable possibility, it is usually by default assume that the local realism theory should reach the same conclusion as the QM for single particle measurement, e.g., both agree with the Malus's law. Suppose there are two particles $A$(lice) and $B$(ob), the general local measurements on them may be expressed as
\begin{align}
\mathcal{M}_{\pm}^{(A)}(\vec{a}) & = \frac{\eta_a^{(\pm)}\pm \alpha_a \vec{\sigma} \cdot \vec{a}}{2} \; , \label{M-local-A}\\
\mathcal{M}_{\pm}^{(B)}(\vec{b}) & = \frac{\eta_b^{(\pm)}\pm \alpha_b \vec{\sigma} \cdot \vec{b}}{2} \; , \label{M-local-B}
\end{align}
where $\eta_{a,b}^{(\pm)}= 1\pm \eta_{a,b}$ and $\alpha_{a,b}$ represent the bias and unsharp parameters on each side, respectively. For a bipartite system composed of $A$ and $B$, let the joint distribution $P_{jk}(\vec{a},\vec{b})$ with $j,k\in \{+,-\}$ being the probability of observing the results $j$ and $k$ on each side by performing the measurements in (\ref{M-local-A}) and (\ref{M-local-B}) along $\vec{a}$ and $\vec{b}$, respectively. Then, we have the following correlation function
\begin{align}
E(\vec{a},\vec{b}\,) \equiv P_{++}(\vec{a},\vec{b}\,) - P_{+-}(\vec{a},\vec{b}\,) -P_{-+}(\vec{a},\vec{b}\,) + P_{--}(\vec{a},\vec{b}\,) \;.
\end{align}
In light of Refs. \cite{CH-general} and \cite{CHSH-unsharp-1} (Lemma 1 of \cite{CH-general} and equation (15) of \cite{CHSH-unsharp-1}), we can put forward two propositions:
\begin{proposition}
In bipartite system, the local realism theory is constrained via joint distributions for biased and unsharp measurements,
\begin{align}
& P_{jk}(\vec{a},\vec{b}) - P_{jk}(\vec{a},\vec{b}')+
P_{jk}(\vec{a}',\vec{b}) +
P_{jk}(\vec{a}',\vec{b}') -   \nonumber \\
& (1+k\eta_b)P_{j}(\vec{a}') - (1+j\eta_a) P_{k}(\vec{b}) + \frac{(1+j\eta_a)(1+k\eta_b)-|\alpha_a\alpha_b|}{2}  \leq 0 \; . \label{CH-POVM}
\end{align}
Here $\eta_{a,b} $ and $\alpha_{a,b}$ are bias and unsharpness parameters on $Alice$ and $Bob$, respectively with $j,k\in \{+1,-1\}$. \label{Proposition-CH}
\end{proposition}
\begin{proposition}
In bipartite system, the local realism theory is constrained via correlation functions for biased and unsharp measurements,
\begin{align}
\left| E(\vec{a},\vec{b})  -  E(\vec{a},\vec{b}') +  E(\vec{a}',\vec{b})  + E(\vec{a}',\vec{b}') \right| \leq 2(|\eta_a|+ |\alpha_a|)(|\eta_b|+|\alpha_b|) \; . \label{CHSH-POVM}
\end{align}
Here $\eta_{a,b}$ and $\alpha_{a,b}$ are bias and unsharpness parameters on $Alice$ and $Bob$, respectively. \label{Proposition-CHSH}
\end{proposition}

\begin{figure}\centering
\scalebox{0.5}{\includegraphics{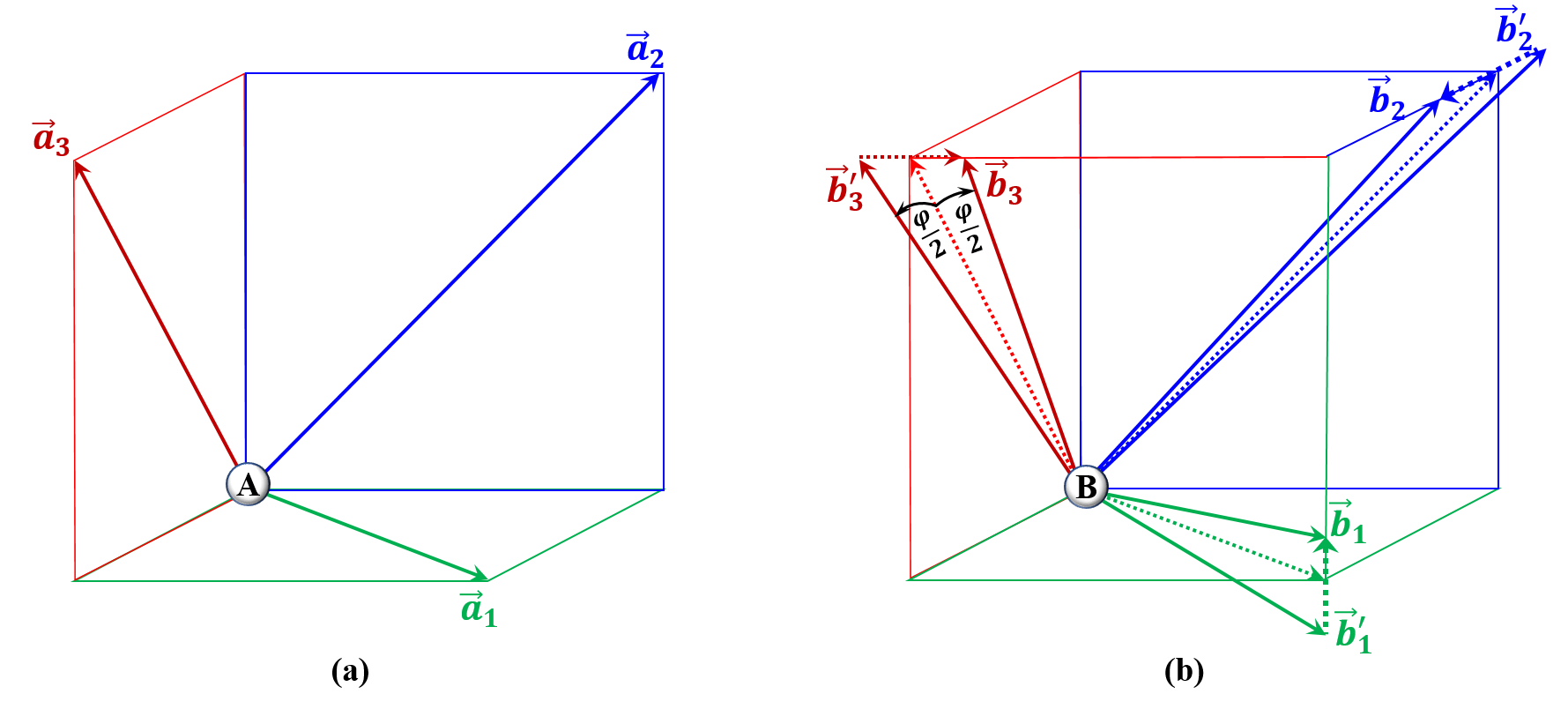}}
\caption{{\bf The triple-measurement settings for the Leggett inequality.} (a) The measurement settings $\vec{a}_i$ are chosen to be in three orthogonal planes for particle $A$. (b) The vectors $\vec{b}_i$ and $\vec{b}_i'$ are so arranged that the three vectors ($\vec{b}_i-\vec{b}_i'$) are orthogonal to each other on particle $B$. Note, $\vec{a}_1$, $\vec{a}_2$, and $\vec{a}_3$ need not to be orthogonal. } \label{Figure-M-setting}
\end{figure}

Next, we explore the effective Leggett constraint, viz inequality, for the general POVM measurements of equations (\ref{M-local-A}) and (\ref{M-local-B}). For two independent particles $A$ and $B$, the local expectation values of given polarizations $|\vec{u}\rangle$ and $|\vec{v}\rangle$ write
\begin{align}
\bar{A}_{\vec{u}} (\vec{a}) & = \eta_a + \alpha_a \vec{u} \cdot \vec{a} \; , \\
\bar{B}_{\vec{v}} (\vec{b}\,) & = \eta_b + \alpha_b \vec{v} \cdot \vec{b} \; .
\end{align}
By means of the procedure proposed in Ref. \cite{Leggett-NP}, we may obtain
\begin{proposition}
In bipartite system, the correlation functions in Leggett nonlocal realism model endure the following constraints for biased and unsharp measurement:
\begin{align}
\frac{1}{3} \sum_{i=1}^3 \left| E(\vec{a}_i,\vec{b}_i) + E(\vec{a}_i,\vec{b}_i') \right| + \frac{2|\alpha_b|}{3}\left|\sin\frac{\varphi}{2}\right|& \leq 2 \; . \label{Leggett-POVM-1}
\end{align}
Here, for each $i$, the unit vectors $\vec{a}_i$ lie in the middle of $\vec{b}_i$ and $\vec{b}_i'$; $\vec{b}_i$ and $\vec{b}_i'$ have the same polar angle $\varphi$ and are also unit, as shown in Figure \ref{Figure-M-setting}. \label{Proposition-Leggett}
\end{proposition}
The derivation of Proposition \ref{Proposition-Leggett} is presented in the Appendix. It is interesting to see that for unbiased measurement the Bell inequality (\ref{CHSH-POVM}) is homogeneous with respect to the unsharp parameters $\alpha_{a,b}$, while the Leggett inequality (\ref{Leggett-POVM-1}) is inhomogeneous. The advent of inhomogeneity may attribute to the fact that the Leggett inequalities also involve individual properties of the bipartite system \cite{Leggett-NP}.

\subsection{Quantum predictions under the POVM measurement}

The quantum predictions for the joint distribution and correlation function can be evaluated by
\begin{align}
& P_{jk}(\vec{a},\vec{b}\,)   = \langle \mathcal{M}_j^{(A)} \otimes \mathcal{M}_k^{(B)} \rangle \; , \; i,j\in\{+,-\} \; , \\
& E(\vec{a},\vec{b}\,)   = \langle (\mathcal{M}_+^{(A)} - \mathcal{M}_-^{(A)} )\otimes ( \mathcal{M}_+^{(B)} - \mathcal{M}_-^{(B)}) \rangle\;.
\end{align}
Then for spin singlet state $|\psi\rangle_{AB} = \frac{1}{\sqrt{2}}(|+-\rangle - |-+\rangle)$ and unit vectors $\vec{a}$ and $\vec{b}$, we have
\begin{align}
P_{jk}(\vec{a},\vec{b}\,) & = \frac{\eta_a^{(j)}\eta_b^{(k)} +jk\alpha_a\alpha_b (\vec{a}\cdot\vec{b})}{4} \; ,  \label{P-eta-alpha} \\
E(\vec{a},\vec{b}) & = \eta_a\eta_b - \alpha_a\alpha_b (\vec{a} \cdot \vec{b}\,) \; . \label{E-eta-alpha}
\end{align}
The contradiction between the local or nonlocal realism with the quantum prediction may exhibit while substituting (\ref{P-eta-alpha}) and (\ref{E-eta-alpha}) into the Propositions \ref{Proposition-CH}-\ref{Proposition-Leggett}. For the sake of simplicity and applying to the hyperon decay, we restrict only to the case of null bias parameters $\eta_a=\eta_b=0$, i.e., the measurements are unbiased.
\begin{corollary}
For the unbiased measurement, equation (\ref{CHSH-POVM}) turns to the following form in the singlet state:
\begin{align}
|\alpha_a\alpha_b|\left|\vec{a}\cdot \vec{b} - \vec{a} \cdot \vec{b}' + \vec{a}'\cdot \vec{b}+\vec{a}'\cdot \vec{b}' \right| \leq 2 |\alpha_a\alpha_b|\; .
\end{align}
That is, the local realism constraint is always violated by QM if the measurements are not totally unsharp, i.e., $|\alpha_a\alpha_b| >0$.
\end{corollary}
\begin{corollary}
For the unbiased measurement, equation (\ref{Leggett-POVM-1}) turns to the following form in the singlet state:
\begin{align}
|\alpha_a\alpha_b|\left|\vec{a}_i\cdot\vec{b}_i+\vec{a}_i\cdot\vec{b}_i'\right| + \frac{2|\alpha_b|}{3}|\sin\frac{\varphi}{2}| \leq 2 \; . \label{Leggett-ab-R}
\end{align}
That is, the nonlocal realism constraint can be violated by the QM, if the measurements are sharp enough, i.e., $(\alpha_a^2 + \frac{1}{9})\cdot \alpha_b^2 > 1$.  \label{Corollary-Leggett}
\end{corollary}

\section{Test the Leggett inequality with entangled hyperon pair under POVM measurement}

\begin{figure}\centering
\scalebox{1}{\includegraphics{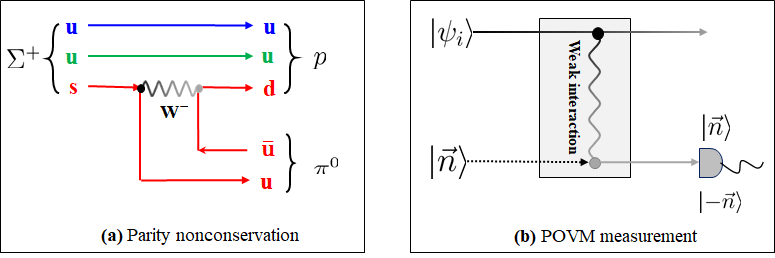}}
\caption{{\bf Parity nonconservation induction of the POVM measurement on hyperon spin.} (a) In polarized $\Sigma^+$ decay, due to the weak interaction, the initial baryon spin projects in the direction of final momentum, and the parity violation may smear the projection. (b) The POVM measurement scenario for baryon weak decay, where unsharpness is described by the asymmetry decay parameter $\alpha$.} \label{Figure-Baryon-decay}
\end{figure}

The POVM measurement, according to the method developed in Ref. \cite{CH-general}, for the hyperon hadronic decay $\Sigma^+\to p\pi^0$ is shown in Figure \ref{Figure-Baryon-decay} for exhibition. Here the Hilbert spaces of spin is coupled to momentum by the weak interaction $U_{\mathrm{w}}$ in the following form:
\begin{align}
U_{\mathrm{w}}: |\psi_i\rangle \otimes |\vec{n}\rangle \mapsto M_{+}(\vec{n})|\psi_i\rangle \otimes |\vec{n} \rangle + M_-(\vec{n})|\psi_i\rangle \otimes |-\vec{n}\rangle \; ,
\end{align}
where $\vec{n}$ is the unit vector of the momentum of the final state hadron, and
\begin{align}
M_{\pm}(\vec{n}) &= \frac{1}{[2(|S|^2+|P|^2)]^{1/2}} (S  \pm P\vec{\sigma} \cdot \vec{n}) \; .
\end{align}
Here $S$ and $P$ are the decay amplitudes in $s$ and $p$ waves. Let $\mathcal{M}_{\pm}(\vec{n}) \equiv M_{\pm}^{\dag}(\vec{n})M_{\pm}(\vec{n})$, and the reduced density matrix for specific momentum reads
\begin{align}
\rho_{n} = \langle \psi_i| \mathcal{M}_{+}(\vec{n}) |\psi_i\rangle \otimes |\vec{n}\rangle \langle \vec{n}| + \langle \psi_i| \mathcal{M}_{-}(\vec{n}) |\psi_i\rangle \otimes |-\vec{n}\rangle \langle -\vec{n}|\; .
\end{align}
Thus, the projective measurement in direction $\vec{n}$ exerts a weak measurement on spin,
\begin{align}
P_+(\vec{n}) & = \langle\psi_i|\mathcal{M}_{+}(\vec{n})|\psi_i\rangle = \left\langle \frac{1 + \alpha \vec{\sigma} \cdot \vec{n}}{2}\right\rangle \; , \\
P_-(\vec{n}) & = \langle\psi_i|\mathcal{M}_{-}(\vec{n})|\psi_i\rangle = \left\langle \frac{1-\alpha\vec{\sigma}\cdot \vec{n}}{2} \right\rangle \; .
\end{align}
Here $\alpha = (S^*P+SP^*)/(|S|^2 + |P|^2)$ is the decay parameter. Comparing to equations (\ref{M-eta-alpha}) and (\ref{Probability-M}), we find that the hyperon two-body hadronic decay behaves as a POVM measurement, which is unbiased and bears the unsharp parameter $\alpha$.

\begin{figure}\centering
\scalebox{0.9}{\includegraphics{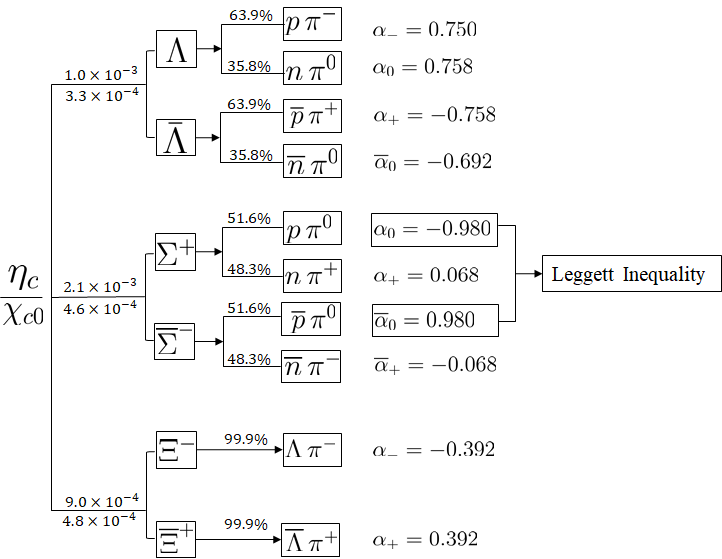}}
\caption{{\bf The decay modes and asymmetric parameters for various hyperon-pair production \cite{PDG}.} The asymmetry parameter $\alpha$ characterize unsharpness of spin measurement in the weak decay. While all parameters can violate the CH and CHSH inequalities, only decay mode $\Sigma^+\to p\pi^0$ and its CP conjugation may violate the Leggett inequality.} \label{Figure-etac-list}
\end{figure}

\begin{figure}\centering
\scalebox{0.37}{\includegraphics{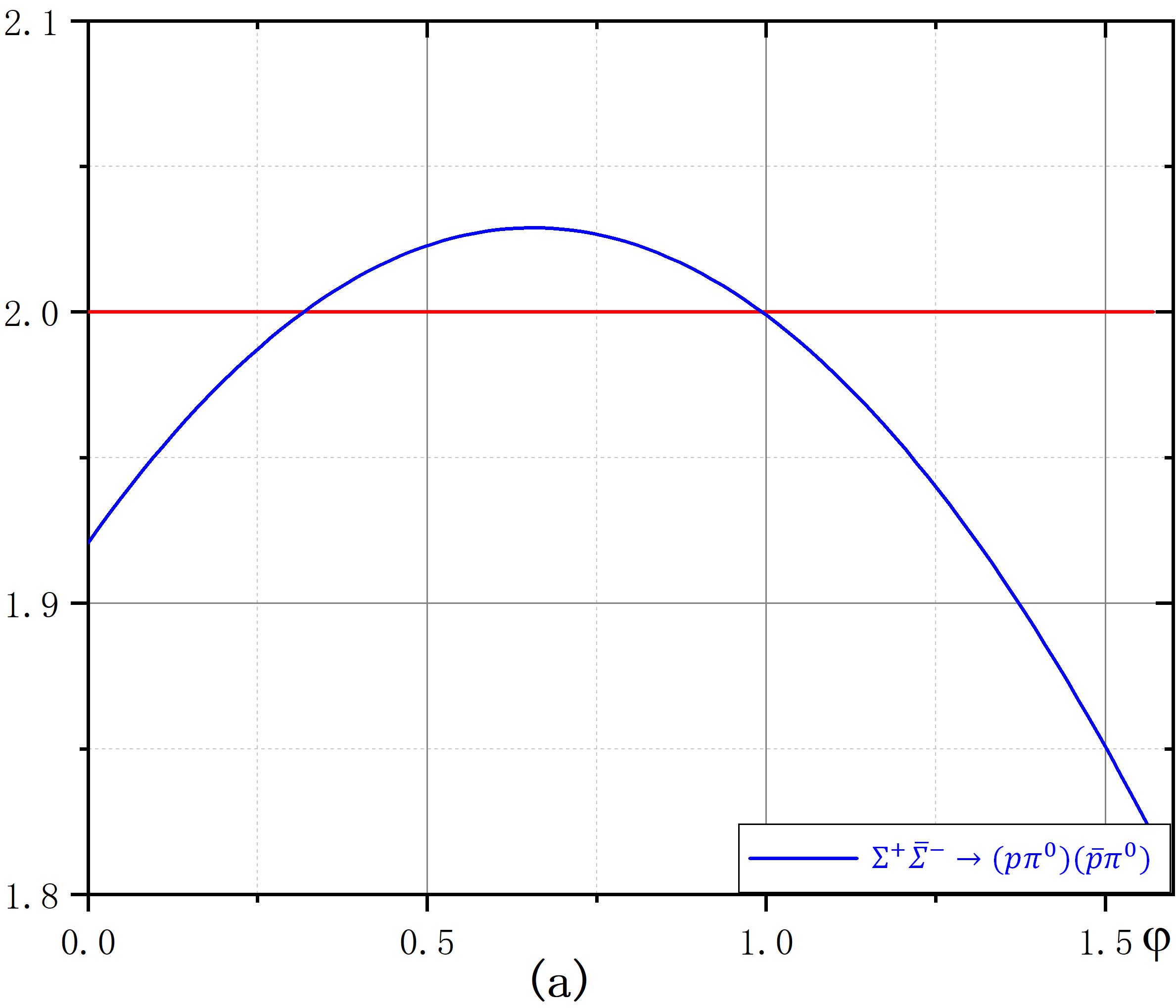}}\;
\scalebox{0.37}{\includegraphics{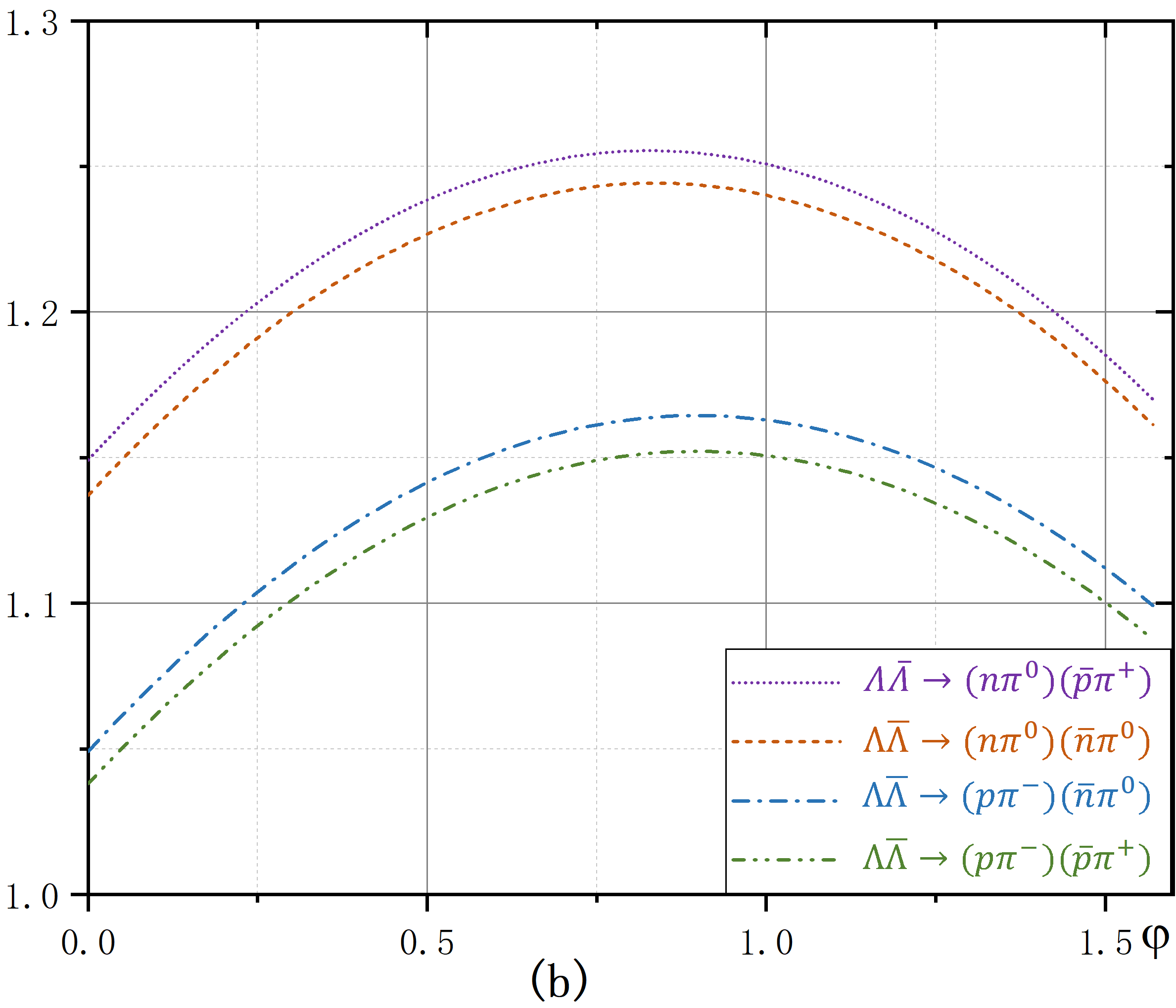}} \\ \vspace{0.3cm}
\scalebox{0.37}{\includegraphics{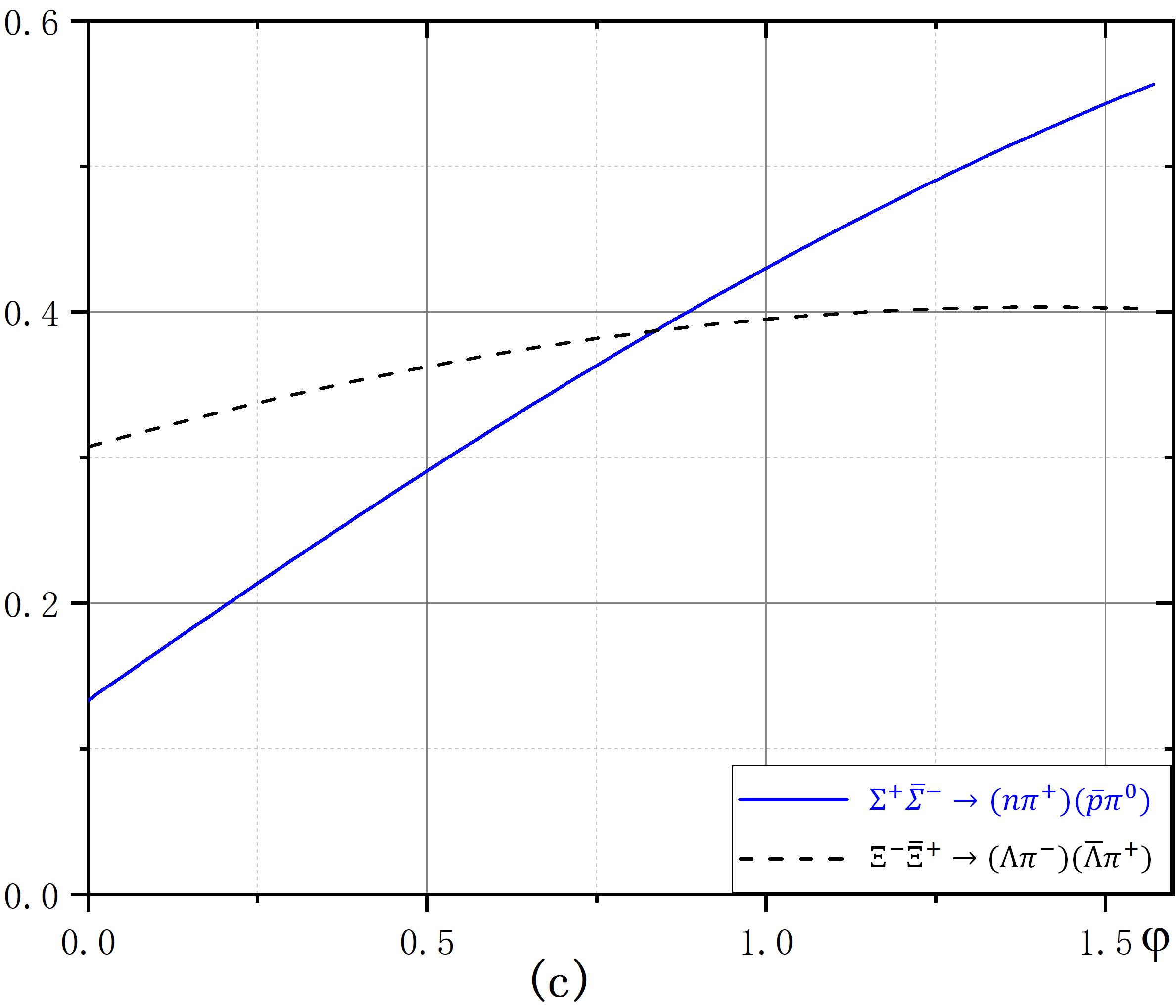}}\;
\scalebox{0.37}{\includegraphics{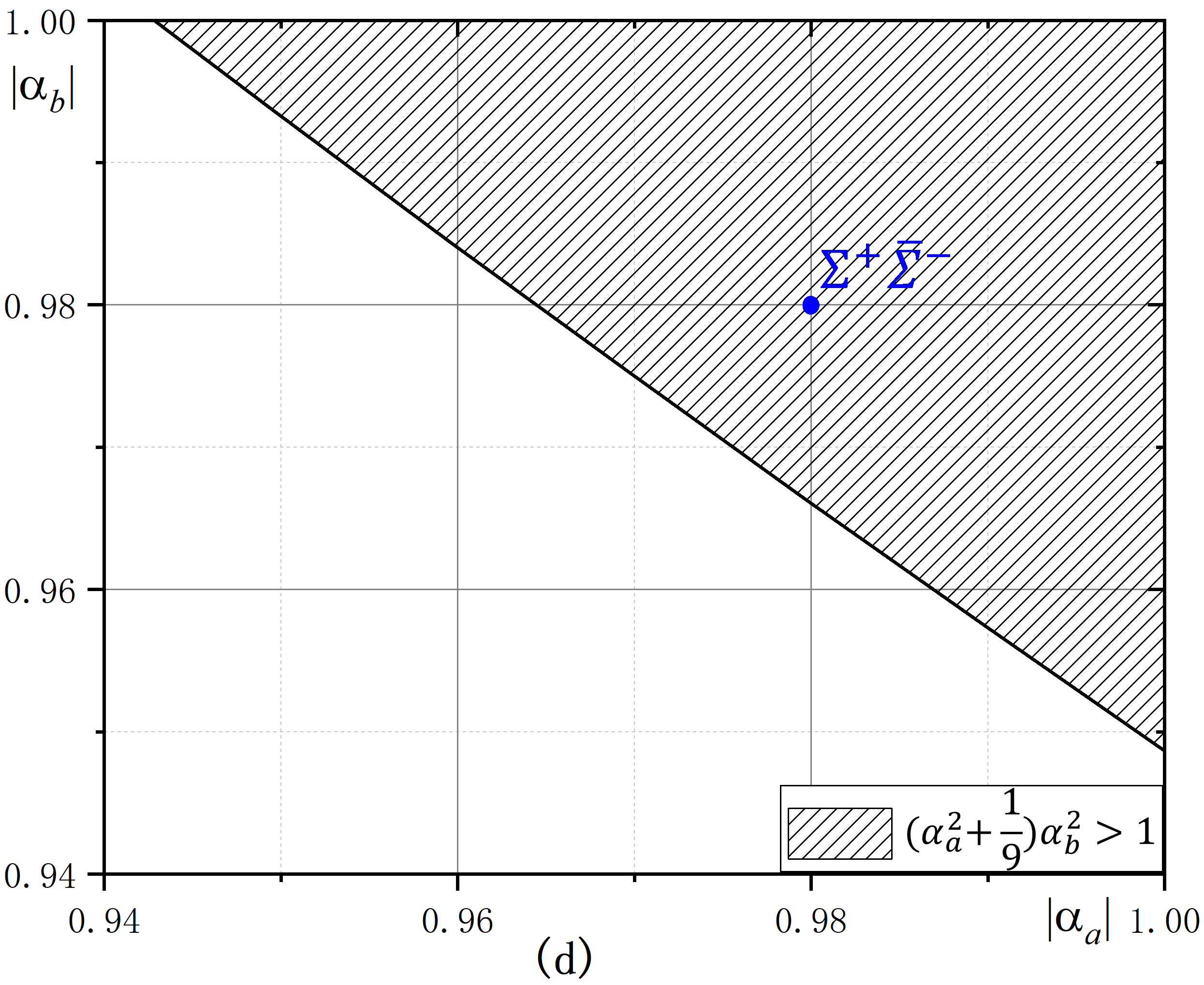}}
\caption{{\bf The violation of the Leggett inequality for entangled hyperon pairs.} For different decay modes, the left-hand side of the Leggett inequalities have fine structures due to the variance of decay parameter $\alpha_{a,b}$, among which the violation we find appears in $\Sigma^+\overline{\Sigma}^- \to (p\pi^0)(\bar{p}\pi^0)$ process. (a)-(c) plot the Leggett inequality (\ref{Leggett-ab-R}) for different decay modes, where the vertical axes signify the magnitudes in Leggett model and horizontal axes for the polar angles of $\vec{b}(\vec{b}')$. A violation of the bound $2$ is found only exist in case (a). The shaded region in (d) exhibits the $\alpha_{a,b}$ regions where the Leggett inequality may be violated.} \label{Figure-Leggett-V}
\end{figure}

In Figure \ref{Figure-etac-list}, various hyperon-pair production processes in $\eta_c$ and $\chi_{c0}$ decays are given, including their decay branching ratios and parameters. The hyperon pairs stemmed from $\eta_c$ are in spin singlet state,
\begin{align}
|\psi\rangle_{AB} = \frac{1}{\sqrt{2}} (|+-\rangle - |-+\rangle) \; . \label{Spin-singlet}
\end{align}
Therefore, by taking the correlation functions and decay parameters into equation (\ref{Leggett-POVM-1}), one may check whether the quantum prediction violates Leggett inequality or not with the POVM measurement induced by weak interaction. In the triple-measurement configuration of Figure \ref{Figure-M-setting}, equation (\ref{Leggett-ab-R}) is plotted for the numerical results of hyperon-pairs production in $\eta_c$ decay, i.e., $\eta_c \to \Sigma^+\overline{\Sigma}^-$,  $\eta_c \to \Lambda \overline{\Lambda}$, and $\eta_c \to \Xi^- \overline{\Xi}^+$ are shown in Figure \ref{Figure-Leggett-V}(a)-(c).
The condition for $\alpha_a$ and $\alpha_b$ on each side where the violation of  equation (\ref{Leggett-ab-R}) could happen is plotted in Figure  \ref{Figure-Leggett-V}(d). The hyperon pairs in $\eta_c$ decay may be symmetric in decay parameter in certain modes, which are noticeable in Figure \ref{Figure-etac-list}. In this case $|\alpha_a|=|\alpha_b| = \alpha$, and Corollary \ref{Corollary-Leggett} gives
\begin{align}
\alpha^4 + \frac{\alpha^2}{9} >1\; .
\end{align}
Hence a violation of Leggett inequality happens only when $\alpha >0.973$.
From the Particle Date Group (PDG) \cite{PDG} we find the channel $\eta_c \to \Sigma^+\overline{\Sigma}^- \to (p\pi^0)(\bar{p}\pi^0)$ fortunately has a enough large $\alpha\sim 0.980$ to break the local realism constraint, see Figures \ref{Figure-etac-list} and \ref{Figure-Leggett-V}(d).


For $\chi_{c0}$ to hyperon-pair channels, we show the Leggett model can be similarly testified. In the rest frame of $\chi_{c0}$, as the most favorite angular momentum of the hyperon pair is $L=1$, the spins of them should fit in the triplet state of $S=1$. The total spin of the hyperon pair should agree with that of $\chi_{c0}$, i.e., $J=0$. Then, the entangled state reads
\begin{align}
|0,0\rangle_{\chi_{c0}} = \frac{1}{\sqrt{3}} (|1,+1\rangle_s|1,-1\rangle_l - |1,0\rangle_s |1,0\rangle_l + |1,-1\rangle_s |1,+1\rangle_l) \; . \label{J0=SL}
\end{align}
Here $s,l$ signify the spin and orbital angular momenta of the hyperon pair. For the angular momentum, the spherical harmonics tell
\begin{align}
\langle \vec{r}\, |1,\pm1\rangle_l\propto
\frac{x\pm iy}{ \sqrt{2}}\; , \; \langle \vec{r}\, |1,0\rangle_l \propto z \; .
\end{align}
If we choose the direction of the outgoing hyperon pair along the $z$ axis, from equation (\ref{J0=SL}) we know the spin state coupled to $|1,0\rangle_l$ takes the following form:
\begin{align}
|\psi\rangle_{AB}= |1,0\rangle_s = \frac{1}{\sqrt{2}}(|+-\rangle + |-+\rangle)\; .
\end{align}
Similar to the spin singlet state (\ref{Spin-singlet}), the correlation function now becomes
\begin{align}
\widetilde{E}(\vec{a},\vec{b}) = \alpha_a\alpha_b ( \vec{\slashed{a}}\cdot\vec{b}\,) \; ,
\end{align}
where $\vec{\slashed{a}} = (a_x,a_y,-a_z) =\mathcal{P}_z \vec{a}$ with $\mathcal{P}_z$ being an inversion along the $z$ axis. To exhibit a correlation as that of the singlet state from $\eta_c$, we need only to perform $\mathcal{P}_z$ to the measurement settings along $A$'s side, i.e.,
\begin{align}
\widetilde{E}(\vec{\slashed{a}},\vec{b}) = \alpha_a\alpha_b(\vec{a}\cdot \vec{b}) = E(\vec{a},\vec{b}) \; . \label{corr-chic}
\end{align}
For the triple-measurement settings in Figure \ref{Figure-M-setting}, the measured quantities become $\widetilde{E}(\vec{\slashed{a}}_i,\vec{b}_i)$ and $\widetilde{E}(\vec{\slashed{a}}_i,\vec{b}_i')$ . With the correlation function (\ref{corr-chic}), all  discussions on $\eta_c$ are applicable to the situation of $\chi_{c0}$, that is to say Figure 4 is also suitable for hyperon pairs coming from $\chi_{c0}$.

\section{Conclusions}

In this work, we generalized the Leggett inequality to incorporate the unsharp POVM measurement. Different from the Bell inequality, the Leggett inequality is found to be inhomogeneous with respect to the unsharpness of the measurement. An unsharp measurement of spin is spontaneously carried out when hyperon undergoes hadronic decays, where the unsharpness arises from the parity violation in the weak interaction. The joint decays of entangled hyperon pair thus can serve as a natural process to test the Leggett model, and we found the violation can be readily observed in the process of $\eta_c(\chi_{c0})\to \Sigma^+\overline{\Sigma}^-\to (p\pi^0)(\bar{p}\pi^0)$ with data accumulated at, for instance, BESIII and BELLE experiments. Because the charge conjugation, parity, and time reversal are all conserved in optical or atomic systems, the experiment test of Leggett inequality in hyperon decays turns out to be a distinctive and indispensable verification for the nonlocal realism. Most importantly, the scheme we propose employing particle decay as the POVM measurement provides an ideal means to the study of nonlocal theories in high energy physics.

\section*{Acknowledgements}
\noindent
This work was supported in part by the Strategic Priority Research Program of the Chinese Academy of Sciences, Grant No.XDB23030100 and by the National Natural Science Foundation of China(NSFC) under the Grants No. 11975236, No. 11635009, and No. 11375200.

\newpage

\setcounter{figure}{0}
\renewcommand{\thefigure}{S\arabic{figure}}
\setcounter{equation}{0}
\renewcommand\theequation{S\arabic{equation}}
\setcounter{theorem}{0}
\renewcommand{\thetheorem}{S\arabic{theorem}}
\setcounter{observation}{0}
\renewcommand{\theobservation}{S\arabic{observation}}
\setcounter{proposition}{0}
\renewcommand{\theproposition}{S\arabic{proposition}}
\setcounter{lemma}{0}
\renewcommand{\thelemma}{S\arabic{lemma}}
\setcounter{corollary}{0}
\renewcommand{\thecorollary}{S\arabic{corollary}}
\setcounter{section}{0}
\renewcommand{\thesection}{S\arabic{section}}

\appendix{\bf \Huge Appendix}

\section{The POVM measurement of spin in quantum field theory}

In quantum mechanics, a general measurement is described by the following: A measurement described by measurement operator $M_{m}$ is performed upon the quantum state $|\psi\rangle$, then the probability of outcome $m$ is given by $P_m = \langle \psi|M_m^{\dag}M_{m}|\psi\rangle$, and the normalization of the probability requires $\sum_m M_{m}^{\dag}M_{m} = \mathds{1}$. This is called the Positive Operator-Valued Measure(POVM) measurements and the operators $\mathcal{M}_m\equiv M_{m}^{\dag}M_m$ are known as the POVM elements \cite{S-QPQI-Book}. Take the qubit state as example, a POVM measurement can be realised by coupling the state with an auxiliary measurement apparatus via some unitary interaction $U_{\mathrm{I}}$ \cite{S-QC-Book}
\begin{align}
U_{\mathrm{I}}: |\psi\rangle \otimes |0\rangle \mapsto  M_+ |\psi\rangle \otimes |+\rangle + M_- |\psi\rangle \otimes |-\rangle \;.
\end{align}
The probability of projecting the apparatus onto the pointer basis $\{|\pm\rangle\}$ gives the measurement of $|\psi\rangle$ with outcomes $\pm$ with probability $P_{\pm} = \langle \psi| M_{\pm}^{\dag}M_{\pm}|\psi\rangle$. It is clear that the projective measurement is returned if we let $M_{m} = |m\rangle\langle m|$, i.e. $\mathcal{M}_m = M_m^{\dag}M_{m} = M_m$, and
\begin{align}
P_{m} = \langle \psi|\mathcal{M}_{m}|\psi\rangle = \langle \psi|m\rangle \langle m|\psi \rangle = |\langle m|\psi\rangle|^2 \; ,
\end{align}
where for qubit system the measurement results may be set as $m=\pm1$.

For hadronic decay of the hyperon with $J^P=\frac{1}{2}^{+}$, the decay amplitude may be expressed as proportional to the following  \cite{S-Weak-ILQ-Book}
\begin{align}
\mathscr{M}   \propto \langle \psi(\vec{n}_f)| (S+P\vec{\sigma} \cdot \vec{n})|\psi(\vec{n}_i)\rangle \; .
\end{align}
Here $|\psi(\vec{n}_{i,f})\rangle$ are the spinors of the initial and final fermions; $S$ and $P$ can be interpreted as the $s$-wave and $p$-wave contributions to the amplitude; $\vec{n} = \vec{p}_f/|\vec{p}_f|$ is the unit vector of final baryon momentum. Summing over the polarizations of the final state baryon, the transition probability is
\begin{align}
\sum_{f\ \mathrm{spins}} |\mathscr{M} |^2 & \propto \sum_{f\ \mathrm{spins}} \mathrm{Tr}\left[| \psi(\vec{n}_f)\rangle \langle \psi(\vec{n}_f)| (S+P\vec{\sigma} \cdot \vec{n})|\psi(\vec{n}_i)\rangle \langle \psi(\vec{n}_i)| (S^* +P^* \vec{\sigma} \cdot \vec{n}) \right] \nonumber \\
& = \mathrm{Tr}\left[ (S+P\vec{\sigma} \cdot \vec{n})|\psi(\vec{n}_i)\rangle \langle \psi(\vec{n}_i)| (S^* +P^* \vec{\sigma} \cdot \vec{n}) \right] \nonumber \\
& = \langle \psi(\vec{n}_i)| (S^* +P^* \vec{\sigma} \cdot \vec{n})(S +P \vec{\sigma} \cdot \vec{n}) |\psi(\vec{n}_i) \rangle \nonumber \\
& = |S|^2 + |P|^2 + (S^*P+SP^*)\vec{n}_i\cdot\vec{n} \propto 1+\alpha\vec{n}_i\cdot\vec{n}\; . \label{S-Amplitude-M}
\end{align}
Here $\alpha = (S^*P+SP^*)/(|S|^2 + |P|^2)$. Based on the method introduced in \cite{S-CH-general}, we may formulate the following POVM measurement model for the above weak interaction process. The Hilbert spaces of the spin of hyperon and the momentum of the final state bayron are coupled by the weak interaction $U_{\mathrm{w}}$ in form of
\begin{align}
U_{\mathrm{w}}: |\psi\rangle \otimes |\vec{n}\rangle \mapsto M_+(\vec{n})|\psi\rangle \otimes |\vec{n}\rangle + M_-(\vec{n})|\psi\rangle \otimes |-\vec{n}\rangle \; ,
\end{align}
where $\vec{n}$ is the unit vector of the momentum of final state baryon and
\begin{align}
M_{\pm}(\vec{n}) \equiv  \displaystyle \frac{1}{\sqrt{2(|S|^2+|P|^2)}} \left[S + P  \vec{\sigma} \cdot (\pm\vec{n}) \right] \; .
\end{align}
We have
\begin{align}
\mathcal{M}_{\pm}(\vec{n}) = M_{\pm}(\vec{n})^{\dag}M_{\pm}(\vec{n}) = \frac{1\pm \alpha\vec{\sigma} \cdot\vec{n}}{2} \; .
\end{align}
This is unbiased and unsharp POVM measurements with $\alpha$ characterizing the unsharpness. The probability for observing $+1$ when measuring along $\vec{n}$ is
\begin{align}
P_{+}(\vec{n}) = \langle \psi(\vec{n}_i)| \mathcal{M}_{+}(\vec{n}) |\psi(\vec{n}_i)\rangle = \frac{1+\alpha\vec{n}_i \cdot\vec{n}}{2} \; . \label{S-POVM-W}
\end{align}
Equation (\ref{S-POVM-W}) is consistent with equation (\ref{S-Amplitude-M}), and it is easy to check the probabilities are normalized $P_+(\vec{n})+P_-(\vec{n}) = 1$.

\section{The Leggett inequality for POVM measurements}

Following the method of Ref. \cite{S-Leggett-Exp1}, we formulate the Leggett's non-local model with the following specific photon source. Suppose the source emits pairs of photons with well-defined polarizations $\vec{u}$ and $\vec{v}$ to $A$ and $B$ respectively. The local measurement outcomes are fully determined as
\begin{align}
\bar{A}_{\vec{u},\vec{v}}(\vec{a}) & = \int  A(\vec{a},\vec{b},\xi) G_{\vec{u}, \vec{v}}(\xi)\, \mathrm{d} \xi = \vec{u}\cdot \vec{a}\;, \label{S-Auv} \\
\bar{B}_{\vec{u},\vec{v}}(\vec{b}) & = \int B(\vec{b},\vec{a},\xi)  G_{\vec{u}, \vec{v}}(\xi)\, \mathrm{d}\xi =\vec{v}\cdot \vec{b}\;. \label{S-Buv}
\end{align}
Here $G_{\vec{u},\vec{v}}(\xi)$ is a normalized distribution describing the subensembles with definite polarization of $\vec{u}$ and $\vec{v}$. (For POVM measurement, we need to replace Malus' law with the corresponding form as in the main text. $\vec{a}$ and $\vec{b}$ are the unit directions in lab where the final state particles are going.) The correlation term is
\begin{align}
C_{\lambda}(\vec{a},\vec{b}) = \int A(\vec{a},\vec{b},\xi) B(\vec{b},\vec{a},\xi) G_{\lambda}(\xi) \, \mathrm{d}\xi \; , \label{S-Cuv}
\end{align}
where $\lambda :=\vec{u}\otimes\vec{v}$. In the spontaneous decays, the correlation may take the form of
\begin{align}
C_{\lambda}(\vec{a},\vec{b}) = \int A\left[\vec{a}(\xi),\vec{b}(\xi),\xi\right] B\left[\vec{b}(\xi),\vec{a}(\xi),\xi\right] G_{\lambda}(\xi) \, \mathrm{d}\xi \; . \label{S-Cuvxi}
\end{align}
That is, the measurement settings are predetermined by the hidden variables. Within this general source producing mixtures of polarized photons, the correlation function for the whole ensemble is given by
\begin{align}
E(\vec{a},\vec{b}) = \int C_{\lambda}(\vec{a},\vec{b}) F(\lambda) \mathrm{d}\lambda = \int C_{\vec{u},\vec{v}}(\vec{a},\vec{b}) F(\vec{u},\vec{v}) \, \mathrm{d}\vec{u}\mathrm{d}\vec{v} \; .
\end{align}
Here $F(\lambda)$ is a distribution function of the source polarizations. For singlet state, there is no prior directions in the lab that the polarizations of the source concentrated to. Hence $F(\vec{u},\vec{v})$ is isotropic with respect to the real space directions.

According to Ref. \cite{S-Leggett-NP}, the joint distribution for observing $a,b\in \{+1,-1\}$ with the measurements along $\vec{a}$ and $\vec{b}$ in a bipartite system may always be written as
\begin{align}
P_{\lambda}(a,b|\vec{a},\vec{b}) = \frac{1}{4} \left[1 + a  M_{\lambda}^{(A)}(\vec{a},\vec{b}) +  b M_{\lambda}^{(B)}(\vec{a},\vec{b})+ ab C_{\lambda}(\vec{a},\vec{b})\right] \; . \label{S-joint-PDF}
\end{align}
From equation (\ref{S-joint-PDF}), it is easy to show that the correlation term is
\begin{align}
C_{\lambda}(\vec{a},\vec{b}) = \sum_{a,b\in\{+1,-1\}} ab P_{\lambda}(a,b|\vec{a},\vec{b}) \;,
\end{align}
and the marginal terms for $A$ and $B$ are
\begin{align}
M_{\lambda}^{(A)} (\vec{a},\vec{b}) & = \sum_{a,b\in\{+1,-1\}} a P_{\lambda}(a,b|\vec{a},\vec{b}) =  \bar{A}_{\lambda}(\vec{a})\; , \label{S-Marginal-A} \\
M_{\lambda}^{(B)}(\vec{a},\vec{b}) & = \sum_{a,b\in\{+1,-1\}} b  P_{\lambda}(a,b|\vec{a},\vec{b}) = \bar{B}_{\lambda}(\vec{b}) \; , \label{S-Marginal-B}
\end{align}
where $\bar{A}_{\lambda}(\vec{a})$, $\bar{B}_{\lambda}(\vec{b})$, and $C_{\lambda}(\vec{a},\vec{b})$ have the same meaning as that of equations (\ref{S-Auv})-(\ref{S-Cuvxi}).

The joint probability distribution should be positive semi-definite, i.e.
\begin{align}
P_{\lambda} (+1,+1|\vec{a},\vec{b} )\geq 0 \; , \; P_{\lambda} (+1,-1|\vec{a},\vec{b} )\geq 0\;,   \\ \; P_{\lambda} (-1,+1|\vec{a},\vec{b} )\geq 0 \; , \; P_{\lambda} (-1,-1|\vec{a},\vec{b} )\geq 0 \; ,
\end{align}
which gives the following inequalities for the marginals and correlations
\begin{eqnarray}
-\left[ 1+ C_{\lambda}(\vec{a},\vec{b}\,)  \right]  \leq &
M_{\lambda}^{(A)} (\vec{a},\vec{b}) +
M_{\lambda}^{(B)} (\vec{a},\vec{b}) & \leq 1 + C_{\lambda}(\vec{a},\vec{b}\,) \; , \\
-\left[ 1- C_{\lambda}(\vec{a},\vec{b}\,)  \right]  \leq &
M_{\lambda}^{(A)} (\vec{a},\vec{b}) -
M_{\lambda}^{(B)} (\vec{a},\vec{b}) & \leq 1 - C_{\lambda}(\vec{a},\vec{b}\,) \; .
\end{eqnarray}
Using equations (\ref{S-Marginal-A}) and (\ref{S-Marginal-B}) to replace $M_{\lambda}^{(A,B)}$ and for two directions on $\vec{b}$ and $\vec{b}'$ on $B$, we have
\begin{eqnarray}
-\left[ 1+ C_{\lambda}(\vec{a},\vec{b}\,)  \right]  \leq & \bar{A}_{\lambda}(\vec{a}\,) +\bar{B}_{\lambda}(\vec{b}\,) & \leq 1 + C_{\lambda}(\vec{a},\vec{b}\,) \; , \label{S-CABC1}\\
-\left[ 1+ C_{\lambda}(\vec{a},\vec{b}'\,)  \right]  \leq & \bar{A}_{\lambda}(\vec{a}\,) +\bar{B}_{\lambda}(\vec{b}'\,) & \leq 1 + C_{\lambda}(\vec{a},\vec{b}'\,) \; ,\label{S-CABC2} \\
-\left[ 1- C_{\lambda}(\vec{a},\vec{b}\,)  \right]  \leq & \bar{A}_{\lambda}(\vec{a}\,) - \bar{B}_{\lambda}(\vec{b}\,) & \leq 1 - C_{\lambda}(\vec{a},\vec{b}\,) \; , \label{S-CABC3}\\
-\left[ 1- C_{\lambda}(\vec{a},\vec{b}'\,)  \right]  \leq & \bar{A}_{\lambda}(\vec{a}\,) - \bar{B}_{\lambda}(\vec{b}'\,) & \leq 1 - C_{\lambda}(\vec{a},\vec{b}'\,) \;. \label{S-CABC4}
\end{eqnarray}
By eliminating $\bar{A}_{\lambda}(\vec{a})$ in equations (\ref{S-CABC1})-(\ref{S-CABC4}), we obtain
\begin{align}
\left| C_{\lambda}(\vec{a},\vec{b}\,) + C_{\lambda}(\vec{a},\vec{b}'\,)\right|  & \leq 2 - \left| \bar{B}_{\lambda}(\vec{b}\,) - \bar{B}_{\lambda}(\vec{b}'\,) \right|\; , \\
\left|C_{\lambda}(\vec{a},\vec{b}\,) - C_{\lambda}(\vec{a},\vec{b}'\,)\right| & \leq 2 - \left| \bar{B}_{\lambda}(\vec{b}\,) + \bar{B}_{\lambda}(\vec{b}'\,)\right| \; .
\end{align}
Integrating over the distribution of the polarization $F(\lambda)$, we get
\begin{align}
\left| E(\vec{a},\vec{b}\,) + E(\vec{a},\vec{b}'\,)\right|  & \leq 2 -|\alpha_b| \int  \left| \vec{v}\cdot (\vec{b} - \vec{b}'\,) \right| F(\lambda)\, \mathrm{d}\lambda\; , \label{S-EEVb-b} \\
\left|E(\vec{a},\vec{b}\,) - E(\vec{a},\vec{b}'\,)\right| & \leq 2 - \int  \left| 2 \eta_b + \alpha_b \vec{v}\cdot(\vec{b}  +  \vec{b}'\,) \right| F(\lambda) \,\mathrm{d}\lambda \; . \label{S-EEVb+b}
\end{align}
Here we have used the followings. First, $\bar{B}_{\lambda} (\vec{b}\,) = \eta_b + \alpha_b \vec{v} \cdot \vec{b}$. Second,
\begin{align}
\int \left|C_{\lambda}(\vec{a},\vec{b})\pm C_{\lambda}(\vec{a},\vec{b}')\right| F(\lambda) \,\mathrm{d}\lambda & \geq \left| \int C_{\lambda}(\vec{a},\vec{b})F(\lambda) \,\mathrm{d}\lambda \pm \int C_{\lambda}(\vec{a},\vec{b}')F(\lambda) \,\mathrm{d}\lambda \right| \nonumber \\
& = \left| E(\vec{a},\vec{b}\,) \pm E(\vec{a},\vec{b}'\,)\right| \; .
\end{align}
Finally, polarization distribution function $F(\vec{u},\vec{v})$ does not vary with the real space directions, i.e. with $\vec{a},\vec{b}$ or $\vec{a}$, $\vec{b}'$.

For triple-measurement setting demonstrated in Figure 1, the equation (\ref{S-EEVb-b}) would give
\begin{align}
\frac{1}{3} \sum_{i=1}^3 \left| E(\vec{a}_i,\vec{b}_i) + E(\vec{a}_i,\vec{b}_i') \right| & \leq 2 -\frac{2|\alpha_b|}{3}\left|\sin\frac{\varphi}{2}\right|\int \left(\sum_{i=1}^3|\vec{v}_i|\right)F(\vec{u},\vec{v})\,\mathrm{d}\vec{u}\mathrm{d}\vec{v}\; .
\end{align}
Because $|v_1|+|v_2|+|v_3|\geq 1$ in orthogonal bases, we have
\begin{align}
\frac{1}{3} \sum_{i=1}^3 \left| E(\vec{a}_i,\vec{b}_i) + E(\vec{a}_i,\vec{b}_i') \right| & \leq 2 - \frac{2|\alpha_b|}{3}\left|\sin\frac{\varphi}{2}\right| \; ,
\end{align}
which is just equation (11). For equation (\ref{S-EEVb+b}),  the integral on the right hand side yields
\begin{align}
\sum_{i=1}^3 \left| 2 \eta_b + \alpha_b \vec{v}\cdot(\vec{b}_i  +  \vec{b}'_i\,) \right| & \geq \sum_{i=1}^3 \left|2|\eta_b| - |\alpha_b| |\vec{v}\cdot (\vec{b}_i+\vec{b}_i')|\right| \nonumber \\
& \geq  \left|6|\eta_b| - 2 |\alpha_b| |\cos\frac{\varphi}{2}| \sum_{i=1}^3 |v_i|\right|\; ,
\end{align}
where we choose the vectors that $(\vec{b}_i+\vec{b}_i') \cdot (\vec{b}_j+\vec{b}_j') = |\vec{b}_i+\vec{b}_i'|\delta_{ij}$. For the unbiased measurement we have
\begin{align}
\frac{1}{3} \sum_{i=1}^3 \left| E(\vec{a}_i,\vec{b}_i) - E(\vec{a}_i,\vec{b}_i') \right| & \leq 2 - \frac{2|\alpha_b|}{3}\left|\cos\frac{\varphi}{2}\right| \;.
\end{align}
Here $1\leq \sum_{i=1}^3|v_i|\leq \sqrt{3}$ and $\vec{b}_i'$ are chosen to be the inverse of those $\vec{b}'_i$ in Figure 1.

\end{document}